# Ad Orientem: the Orientation of Gothic Cathedrals of France

**Amelia Carolina Sparavigna**
Department of Applied Science and Technology
Politecnico di Torino, C.so Duca degli Abruzzi 24, Torino, Italy

*Here the orientation of the Gothic cathedrals in France is discussed and investigated using the satellite maps. Except a few of them, these buildings have the apse facing the rising sun, according to a practice adopted during the middle age.*

The Latin expression "ad orientem" means eastwards. In the catholic liturgy, it describes an eastward orientation of celebrating Mass, according to the "cosmic sign of the rising sun which symbolizes the universality of God." [1,2] It is interesting to note that the earliest churches in Rome had the main entrance facing east and an apse with the altar to the west; the priest celebrating Mass stood behind the altar, facing east and so towards the people. According to Helen Dietz [3], after the Christians in the Rome of the fourth-century freely begin to build churches, they located the sanctuary towards the west end of the building, with the symbolic proposal of imitating the sanctuary of the Jerusalem Temple. It was during the 8th or 9th century that the position changed, and the priest began to face the apse, not the people, when this worship position in celebrating Mass was adopted in the Roman Basilicas. [4] Present-day Roman Missal prefers the priest facing people, but it does not forbid the "ad orientem" position for the priest when saying Mass, only requires that in new or renovated churches the facing-the-people orientation be made possible.

U. Michael Lang writes in [5] that for several religions, the position of worshippers and of worship buildings has a direction according to a "sacred orientation". Believing in a second glorious coming of Jesus Christ, the Christians saw the sunrise as a symbol of resurrection and therefore oriented the sanctuaries Eastwards. Here we will consider the orientation of the gothic cathedrals in France. We will see that an eastward orientation of the apse is a rule. Let us remember that eastwards does not means "due east", therefore it is better to tell that the orientation of cathedrals is towards the direction of the sunrise.

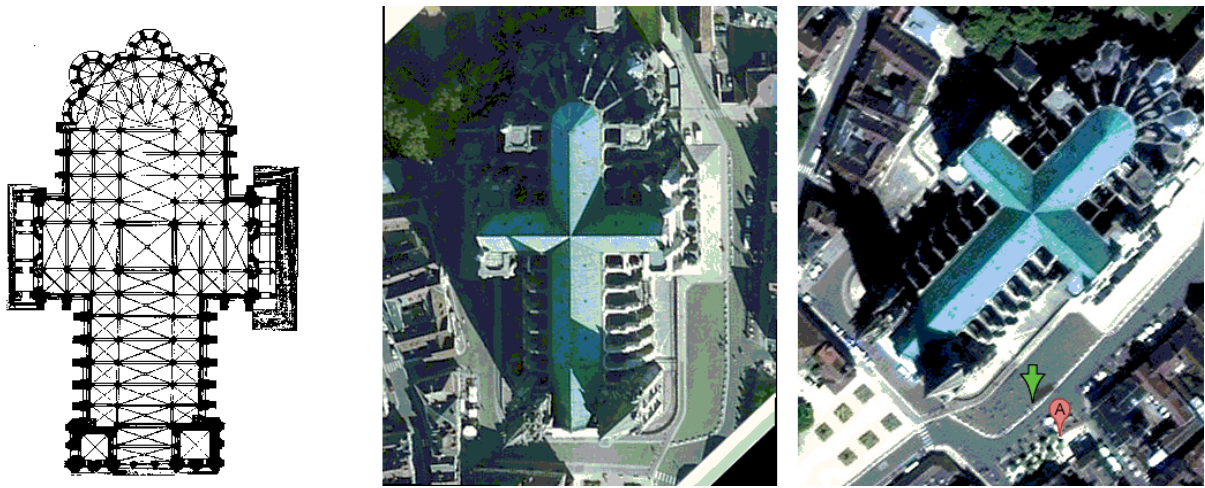

**Fig.1: On the left we can see the floor plan of the Chartres cathedral, shown in the satellite images in the middle and on the right. Note the cruciform structure. In the middle the image is rotated; on the right we see the orientation of the cathedral as it is (Google Maps).**

Before the specific discussion of orientation, let us shortly remember some facts about the plan of gothic cathedrals. A gothic cathedral has usually a cruciform plan, given by an apse, a transept and a nave (see Fig.1). The apse was first a semicircular volume covered with a hemispherical vault. In the gothic cathedral, this term applies to a semi-circular or polygonal termination of the building at the liturgical east end, where the main altar is placed, regardless of the shape of the roof. From the beginning of the XIII Century in France, the apses were built as radiating chapels, as we can see in the Figure 2, where the cathedral of Amiens is shown.

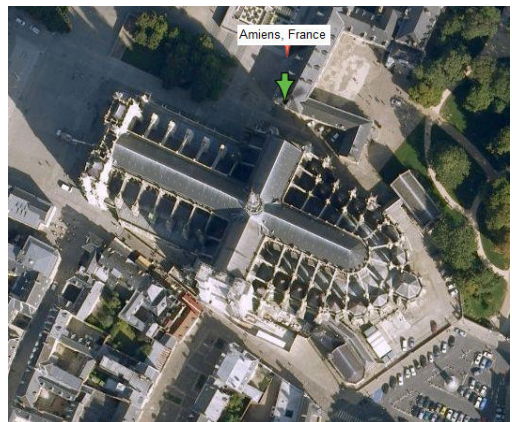

**Fig2. Radiating chapels of the Amiens cathedral (from the Google Maps).**

The transept is a transverse section placed across the main body of a building. In Christian churches having a cruciform plan, the transept is the part separating the nave from the presbytery and the apse. Assuming the altar located at the east end of a church, a transept extends to the north and south. When the transept exceeds the sides of the building, the plan forms the shape of a cross, which can be a "Latin cross" or "Greek cross", in the case that all four extensions have the same length. The "Latin cross" is the preferred plan of gothic cathedrals in France.

The gothic art and architecture in France began about 1140, due to the works of architects and abbots. One of them was the Abbot Suger (1081-1151). Autobiographical accounts are coming from the two books he wrote [6,7]. Suger remodeled the Abbey Church of Saint-Denis, in the northern suburbs of Paris, to achieve some new perspectives. Inspired by the theology of Dionysius, the Syrian Pseudo-Areopagite (ca. 500), Suger saw the universe consisting of the "Father of Lights" (God), the "first radiance" (Christ) and the "smaller lights" (the people) [8]. The church therefore needs to be closer to a universe of light, and this is reflected in Suger's use of heightened architecture and large windows. Reference [8] is telling that the west façade of Saint-Denis cathedral serves as "a stepping-stone on the way to Heaven towards the light of God". The twelve columns in the choir symbolize the twelve apostles, while the columns in the ambulatory the twelve prophets.

According to this highlight of light, a "sacred orientation" of the gothic cathedral towards the rising of the sun is expected. To investigate this fact we can use the satellite maps (Google or Bing Maps) to measure the angle these building are forming with the cardinal West-East direction, assumed as a polar axis. Angles are positive or negative when they are counterclockwise or clockwise with respect to this axis. The measure of the angle from the satellite images is quite easy: it is immediately given by the direction of the roof, which is quite visible (see for instance the case of the cathedral of Amiens, in Fig.3). Of course a check of the this angle on the local horizon could give more precise values. From the satellite, we assume an uncertainty of the measure of ½°. After Figure 3, a table listing cathedrals and corresponding angles is given.

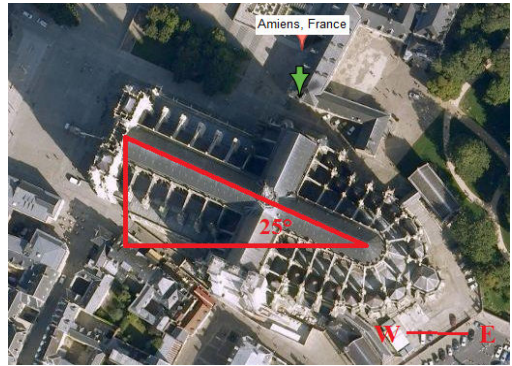

**Fig.3** The West-East direction is giving the polar axis. Angles are positive or negative when they are counterclockwise or clockwise. In the case of the Amiens cathedral the angle is of 25 ° negative.

Table I

| |
|---|
| Metz, Saint-Étienne, 1220, 49° positive |
| Chartres, Notre-Dame, 1194, 46.5° positive |
| Troyes, Saint-Pierre-et-Saint-Paul de Troyes, 1228, 35.5° positive |
| Limoges, Saint-Étienne, 1273, 35° positive |
| Rheims, Notre-Dame, 1211, 31° positive |
| Strasbourg, Notre-Dame, 1225, 29.5° positive |
| Nevers, Saint-Cyr-et-Sainte-Julitte, 1212, 26.5° positive |
| Chalons-en-Champagne, Saint-Étienne, 1230, 17° positive |
| Tours, Saint-Gatien, 1236, 17° positive |
| Evreux, Notre-Dame, 1250, 16° positive |
| Auxerre, Saint-Étienne, 1215, 11.5° positive |
| Bordeaux, Saint-André, 1250, 9.5° positive |
| Toulouse, Saint-Étienne, 1273, 9° positive |
| Sens, Saint-Étienne. 1143, 4.5° positive |
| Langres, Saint-Mammès, 1150, 4° positive |
| Soissons, Saint-Gervais-et-Saint-Protais), 1176, 2.5° positive |
| Meaux, Saint-Étienne, 1200, 0° null |
| Carcassonne, Saint-Nazaire, 1267, 2.5° negative |
| Clermont-Ferrand, Notre-Dame-de-l'Assomption, 1248, 3.5° negative |
| Laon, Notre-Dame, 1150-1156, 3.5° negative |
| Beauvais, Saint-Pierre, 1225, 14.5° negative |
| Bourges, Saint-Étienne, 1195, 19° negative |
| Saint-Denis, Abbey of Saint Denis, 1135-1136, 19° negative |
| Rodez, Notre-Dame, 1277, 19° negative |
| Narbonne, Saint-Just, 1286, 21° negative |
| Bayonne, Sainte-Marie, 1258, 21.5° negative |
| Senlis, Notre-Dame, 1151-1153, 22° negative |
| Paris, Notre-Dame, 1163, 25° negative |
| Amiens, Notre-Dame, 1220, 25° negative |
| Rouen, Notre-Dame-de-l'Assomption,1200, 26° negative |
| Bayeux, Notre-Dame, 1180, 27° negative |
| Noyon, Notre-Dame, 1150, 28° negative |
| Le Mans, Saint-Julien, 1217, 54° negative |

List of Gothic cathedrals after  http://it.wikipedia.org/wiki/Cattedrali_gotiche_francesi

Such an approach using satellite images has been proposed by the author is some papers on the orientation of ancient Egyptian temples and Roman towns [9-11]. At a first glance, some angles of Table I seem quite large, but when we compare them with the sunrise largest amplitudes for the latitudes of France (the red lines are corresponding to the North and South of France, in the diagram of Fig.4), we discover that only three buildings (Metz, Chartres and Le Mans) are exceeding the sunrise orientation.

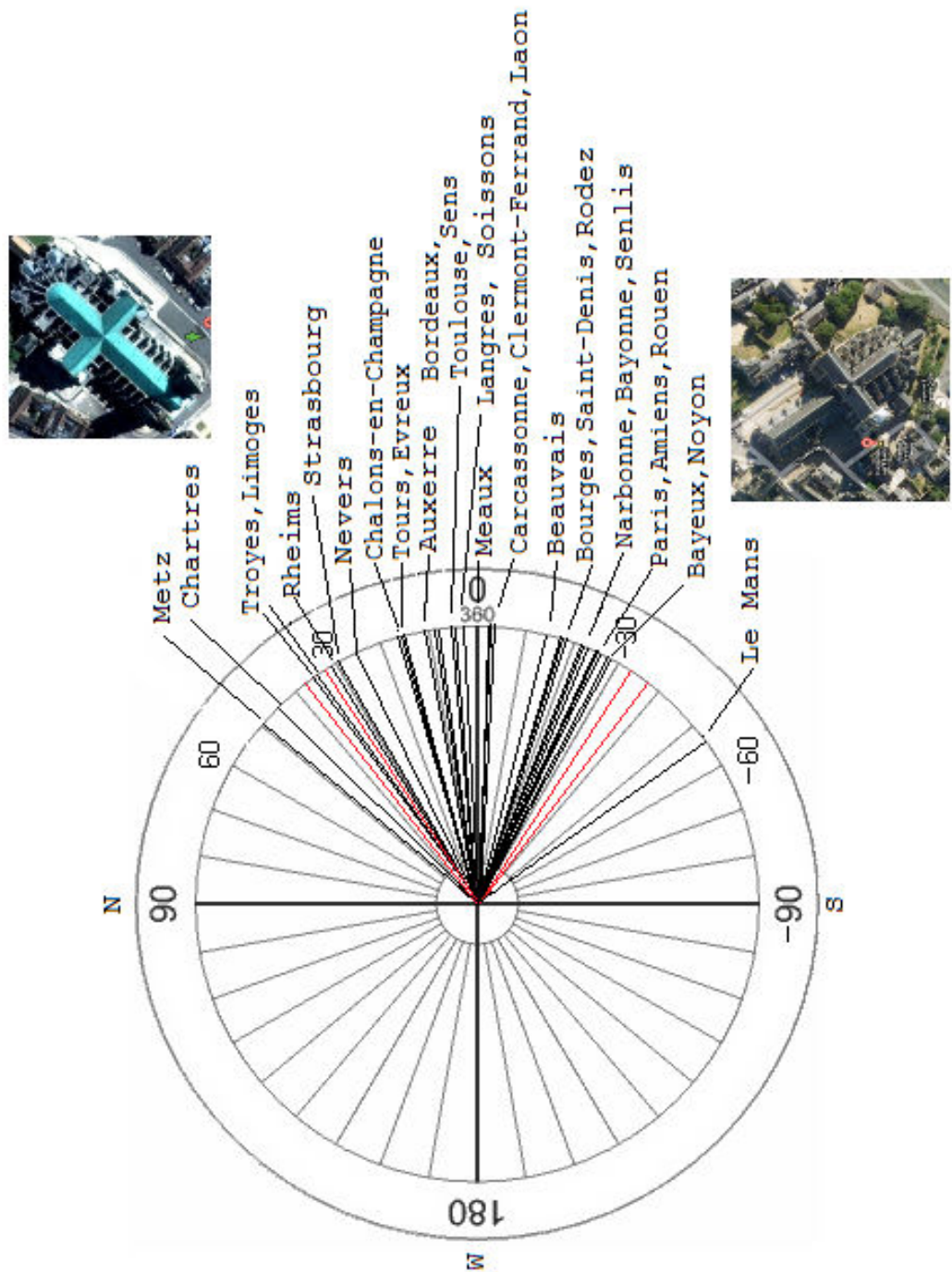

**Fig.4 The angles of the Gothic cathedrals in France on a polar diagram. The red lines correspond to the largest sunrise amplitude for the North and South of France. We see that only three building are exceeding an orientation with the sunrise. They are at Metz, Chartres, and Le Mans.**

In the diagram of Fig.4, a positive angle corresponds to the sunrise amplitude on a day of spring or summer, then from the spring equinox to the autumn equinox, whereas a negative angle corresponds to a day of autumn or winter. A null angle corresponds to an equinox.

The practice of the foundation of a church and its rite had been simplified in 1961, as told in Ref.12. In this document, the orientation of the building is not mentioned. A pre-Vatican II rite dated to the fifteenth century, required, before the construction of a new church, that the foundations of the building were marked out and a wooden cross placed where the altar had to be. Then the bishop blessed the foundation stone. [13,14] This rite too is not mentioning an orientation towards the sunrise. But, as we can read in Ref.15, the Apostolic Constitutions, [16] dated from the fourth century, prescribe that the shape of the church should be oblong, so that it would resemble a ship, that is *Aedes sit oblonga, ad orientem versa, et quae sit navi similis*. Moreover, as reported in [15], the foundation begun on the day, or on the vigil of the feast of the Saint in whose name the new building was to be dedicated. On the night preceding the foundation rite, the "bishop and people would watch upon the site of the future church, and, as soon as the first rays of the sun appeared upon the horizon, the direction or orientation of the church was so determined that the building faced towards that point of the compass. If this theory be true, it would enable us to find out the original dedication (or titular) of a church, where the lapse of time has caused it to be changed or forgotten." This rite, as described by the Benedictine monks, seems the ritual that the ancient Romans used for the foundations of their "castra" and towns [9].

Probably, a large part of the gothic cathedrals listed in Table I and here discussed was built on the sites of more ancient worship places, maintaining their original orientations. In fact we see from the Table I that there are several cathedrals dedicated to Saint Étienne that have quite different angles with respect the cardinal East-West direction. According to [15], this means that the original dedication of the site was different. For what concerns the cathedral of Saint Denis, its orientation from the satellite maps is 19° negative: considering the latitude, this angle corresponds to the sunrise azimuth angle (see [9-11] for definitions) of about 20 October or 20 February. The feast of Saint Denis, celebrated since at least the year 800, was added to the Roman Calendar in the year 1568, and settled for celebration on October 9 [17]. There is a difference of ten days then, perhaps due to the reform to the Gregorian calendar.

Let us conclude with a short remark on the Chartres cathedral: this is one of the three cathedrals among those of the list having an orientation outside the possible sunrise amplitudes as defined in Refs.9-11. Chartres is located on a hill and the cathedral is on the top of this hill. For this reason, the visible horizon is larger and then a correction of the azimuth seems necessary, such as the effect of atmospheric refraction. In the case that corrections are not enough to have a sunrise orientation, let us add this "mystery" to the list of mysteries of this amazing cathedral [18].